\documentclass{ifacconf}

\usepackage{graphicx}      
\usepackage{natbib}        
\usepackage{amsmath}
\usepackage{amsfonts}
\usepackage{amssymb}
\usepackage[labelformat=simple]{subcaption}

\usepackage{bm}

\begin{document}
\begin{frontmatter}

\title{On simplification of Dual-Youla approach for closed-loop identification
\thanksref{footnoteinfo}} 

\thanks[footnoteinfo]{This work has been submitted to IFAC for possible publication. This paper is partly supported by JSPS KAKENHI No. JP17H03281.}

\author[First]{Toshiharu Sugie} 
\author[Second]{Ichiro Maruta} 

\address[First]{Graduate School of Engineering, Osaka University, Suita, Japan, (e-mail: sugie@jrl.eng.osaka-u.ac.jp)}
\address[Second]{Graduate School of Engineering, Kyoto University, Kyoto, Japan, (e-mail: maruta@kuaero.kyoto-u.ac.jp)}


\begin{abstract}                
The dual Youla method for closed loop identification is known to have several practically important merits. Namely, it provides an accurate plant model irrespective of noise models, and fits inherently to handle unstable plants by using coprime factorization. In addition, the method is empirically robust against the uncertainty of the controller knowledge. However, use of coprime factorization may cause a big barrier against industrial applications. This paper shows how to derive a simplified version of the method which identifies the plant itself without coprime factorization, while enjoying all the merits of the dual Youla method. This simplified version turns out to be identical to the stabilized prediction error method which was proposed by the authors recently. Detailed simulation results are given to demonstrate the above merits.
\end{abstract}


\end{frontmatter}

\section{Introduction}

Closed loop identification is often inevitable in real world. When the plant to be estimated is unstable in open loop, the I/O data should be collected in closed loop setting. Even if it is stable, it is often the case that the plant must be operated in the presence of feedback controller due to safety and economic reasons. Since it is well known that closed loop identification is difficult due to the correlation between the measurement noise and the input, various methods have been proposed so far to overcome the difficulty (see e.g., \cite{VandenHof:1995, Forssell:1999, Veen:2013}).

These methods may be classified as direct and indirect ones. Direct methods ignore the presence of feedback controller, and identify the plant model using the I/O data only. However, they require the exact knowledge of noise model structure. It is not easy to obtain such knowledge in most cases. Instead, the knowledge of feedback controller is often available.  In this case, it would be reasonable to exploit the knowledge. Indirect methods use the exact controller knowledge to obtain an accurate plant model subject to noise under modeling. Most of them identify the closed loop transfer function first, then the plant model is calculated. In order to obtain accurate plant model, the exact knowledge of the controller is necessary. Furthermore, it requires some techniques to identify unstable plants as pointed out by \cite{Forssell:2000}. Though this fact may not be well recognized, this could be a serious problem in some cases. 

Among various indirect methods, we focus on the dual Youla method (\cite{Hansen:1988,Hansen:1989, VandenHof:1996}). The method transforms the original closed loop identification into an open loop identification for a stable system by using Youla parametrization based on coprime factorization. This method is inherently robust in identifying {\it unstable} plants, irrespective of the noise models. Moreover, it is {\it not} sensitive to the accuracy of the controller knowledge, which is very important in practice. There are few methods which enjoy both of these merits, except \cite{Aguero:2011}. Unfortunately, the dual-Youla method has some problems to overcome. First, it relies on coprime factorization over a proper stable rational ring $RH_{\infty}$ (see \cite{Vidyasagar:2011}). Since most engineers in industry are not familiar with such coprime factorization, this could be a big barrier for them to use the dual-Youla method. Also, the identified plant model tends to be of high order because of Youla parametrization (unless adopting some technique like the tailor-made parametrization proposed by \cite{VanDonkelaar:2000}). Second, the method tries to identify a virtual system (so called Youla parameter) instead of the plant itself. This is not transparent at all conceptually. Furthermore, it is difficult to exploit prior information of the plant (e.g, integrator type, system order) even if it is available.  

The purpose of this paper is to derive a simplified  identification method which overcomes the above drawbacks based on the dual Youla method for MIMO systems. More precisely, starting from the dual Youla method, we show how to identify the plant itself without coprime factorization.This method turns out to be nothing but the stabilized PEM (prediction error method) developed by  \cite{Maruta:2018}. Furthermore, we will demonstrate the merit to identify 
the plant itself without coprime factorization, and the robustness against the uncertainty of the controller knowledge and noise model structure through detailed simulation.

\section{Dual Youla method}

\begin{figure}[hbt]
 \begin{center}
  \includegraphics[width=0.63\linewidth]{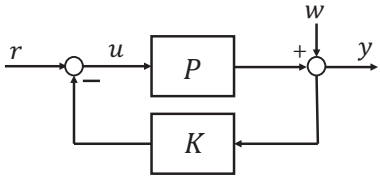} 
\end{center}
\caption{Closed loop system}
\label{fig:1}
\end{figure}

First, we  briefly describe the dual Youla method. Consider the closed loop system shown in Fig \ref{fig:1}, which is described by
\begin{align}
y & = P u + w  
\label{eq:1.1} \\
u &= r -K y   
\label{eq:1.2}
\end{align}
where $P$ is the plant to be identified with $p$-dimensional output $y$ and 
$m$-dimensional input $u$ which may be unstable, and $y$ is contaminated by noise $w$ which could be colored.  
$K$ is a given stabilizing controller, and $r$ is an $m$-dimensional exogenous signal. 
We will identify $P$ based on the information of the I/O data $(u, y)$ with the knowledge of $K$. 

\begin{figure}[hbt]
\centering
\includegraphics[width=0.83\linewidth]{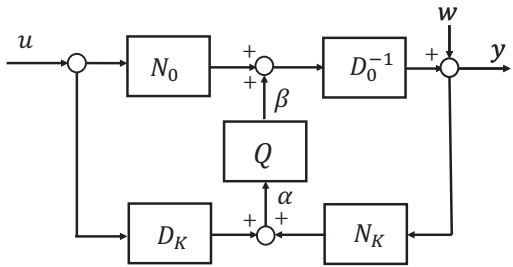} 
\caption{Dual Youla representation of $P$}
\label{fig:2}
\end{figure}

Let $D_K^{-1} N_{K}$ and $\tilde{N}_K \tilde{D}_K^{-1}$ be lcf (left coprime factorization) and rcf
(right coprime factorization) of $K$ over the proper stable rational ring $RH_{\infty}$, respectively. Since $P$ is stabilized by $K$, 
\begin{align}
      P= (D_0 - Q N_K)^{-1}(N_0 + Q D_K) 
\label{eq:1.3}
\end{align}
holds for some $Q \in RH_{\infty}^{p \times p}$, 
where $N_0 \in RH_{\infty}^{p \times m}$ and $D_0 \in RH_{\infty}^{p \times p}$ satisfy the Bézout identity
\begin{align}
 D_0 \tilde{D}_K + N_0 \tilde{N}_K = I_{p}.  
\end{align}
Hence the plant $P$ can be represented by the block diagram shown in Fig. \ref{fig:2}.

\begin{figure}[hbt]
 \begin{center}
  \includegraphics[width=0.43\linewidth]{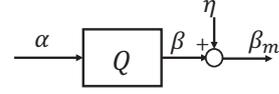} 
\end{center}
\caption{Open loop identification for stable $Q$}
\label{fig:3}
\end{figure}

Let $\alpha$ and $\beta$ be the input and output of  $Q$, respectively.
Then it is easy to see that  
\begin{align}
\alpha &= D_K u + N_K y 
\label{eq:alpha}\\
\beta &= D_0(y-w)-N_0u
 \label{eq:beta}
\end{align}
hold. Let $\beta_m =D_0 y -N_0 u$, then we have
\begin{align}
\beta_m = Q \alpha + \eta,  ~~~ \eta :=D_0 w
\end{align}
Here both $\alpha$ and $\beta_m$ can be calculated from I/O data $(u, y)$.
Moreover, 
$\alpha$ is not correlated with noise $w$ at all because 
\begin{align}
D_K^{-1}\alpha =u-Ky=r
\label{eq:8}
\end{align}
holds from
(\ref{eq:1.2}) and (\ref{eq:alpha}). 
This is very important. 
Hence we can obtain consistent model of $Q$ from $\alpha$ and $\beta$ 
in an open loop setting as shown in Fig. \ref{fig:3}.  
Consequently, we can identify the (possibly unstable) plant $P$ in closed loop by 
using the conventional open loop identification technique.  

{\bf Remark 1:}~ 
The dual Youla method in MIMO case can be found in \cite{VandenHof:1996, Forssell:1999}. However, they use rcf of $K$. So the block-diagram representation is different from the original one in \cite{Hansen:1988}. By using lcf representation of $K$, 
Fig. \ref{fig:2} is the same as Hansen’s one. This enables us to argue the simplification of the dual Youla method as shown below.

{\bf Remark 2:}~ 
Let $\hat{K}$ be any controller which stabilizes $P$, and $D_K^{-1} N_{K}$ and $\tilde{N}_K \tilde{D}_K^{-1}$ be lcf and rcf of $\hat{K}$, respectively. 
Then the above argument still holds. Hence, the original closed loop identification is transformed to an open loop identification of a stable $Q$.
However, in this case, (\ref{eq:8}) should be replaced by
\begin{align}
D_K^{-1}\alpha =u-\hat{K}y= r + (K - \hat{K}) y.
\nonumber
\end{align}
This implies that $\alpha$ is correlated with noise $w$ in case of $K \neq \hat{K}$. As a result, the accuracy of the model $P$ may be degraded
due to this correlation. However, the effect of $K \neq \hat{K}$ is not so direct as the conventional indirect methods.
This might give an intuitive reason why the dual Youla method is {\it not} so sensitive with respect to the controller accuracy.
The numerical example given later supports this insensitivity.

\section{Simplification without coprime factorization}

The above idea is very nice. Furthermore, it is known that the method works well even if the controller knowledge is not so accurate. However, there are some drawbacks. Since the model of $P$ is calculated from (\ref{eq:1.3}) with $Q$, the plant order tends to be higher.  Though there are infinitely many ways of coprime factorization, how to choose them is not clear. 
Above all, most engineers are not familiar with coprime factorization over the proper stable rational ring. So,
it is not easy for them to use this method.  From these observations, we consider
a way to identify the plant itself directly in the same spirit of the dual Youla method in what follows.

\begin{figure}[hbt]
 \begin{center}
  \includegraphics[width=0.93\linewidth]{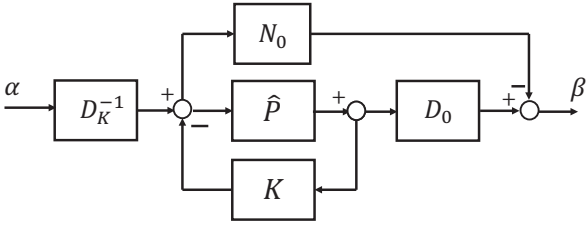} 
\end{center}
\caption{Representation of $\hat{Q}$ in terms of $\hat{P}$ with input $\alpha$ and output $\beta$}
\label{fig:4}
\end{figure}

\begin{figure}[hbt]
 \begin{center}
  \includegraphics[width=0.83\linewidth]{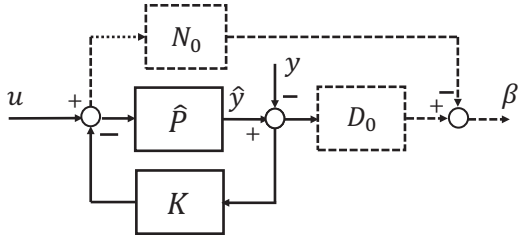} 
\end{center}
\caption{Equivalent representation with input $(u,y)$}
\label{fig:5}
\end{figure}

Let $\hat{Q}$ be the identified model and $\hat{P}$ be the corresponding plant model, namely
\begin{align}
\hat{P}=(D_0 - \hat{Q} N_K)^{-1}(N_0 + \hat{Q} D_K) .
\end{align}
Then we have
\begin{align}
(D_0 -\hat{Q} N_K) \hat{P} &=N_0 + \hat{Q} D_K \nonumber \\
D_0 \hat{P} - N_0  &= \hat{Q} (N_K \hat{P}+ D_K)  \nonumber \\
\hat{Q} &= (D_0 \hat{P} - N_0)(N_K \hat{P} + D_K)^{-1}  \nonumber \\
    &= (D_0 \hat{P}- N_0)(K\hat{P} + I )^{-1}D_K^{-1}
\end{align}
This corresponds to the block diagram shown in Fig. \ref{fig:4}.
Since  $D_K^{-1}\alpha =u-Ky$ hold, the bock-diagram can be transformed into Fig. \ref{fig:5}.

\begin{figure}[hbt]
 \begin{center}
  \includegraphics[width=0.63\linewidth]{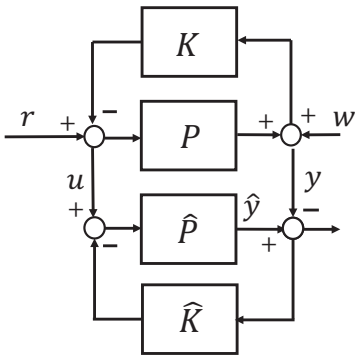} 
\end{center}
\caption{Stabilized Prediction Error Method}
\label{fig:6}
\end{figure}

Now the closed loop part consisting of $(\hat{P}, K)$ is stable and the input of $\hat{P}$ is
not correlated with noise $w$ as in the dual Youla method, we can identify the model 
$\hat{P}$ from $(u,y)$ by minimizing,
e.g., 
\begin{align}
J= \| y - \hat{y} \|_2. \label{eq:J}
\end{align}
This turns out to be nothing but the stabilized PEM (in the case of $\hat{K} = K$) introduced by \cite{Maruta:2018} which is shown in Fig. \ref{fig:6}.

 \section{Numerical examples}
 
\begin{figure}[htbp]
  \centering
  \includegraphics[scale=0.8]{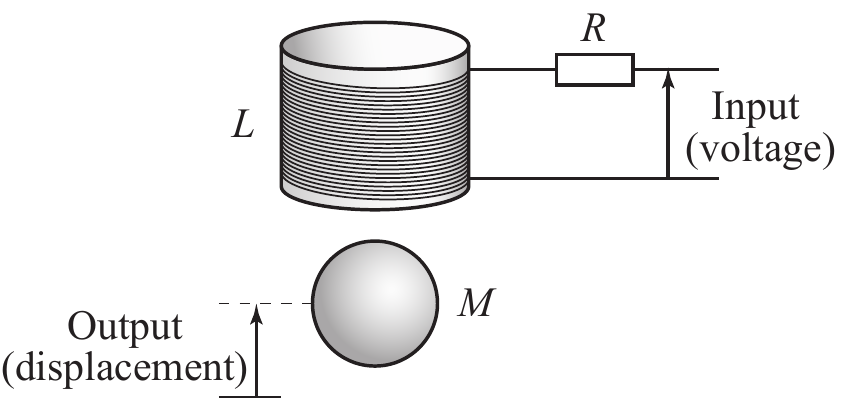}
   \caption{A magnetic levitation system}
  \label{fig:maglev}
\end{figure}
  
We will demonstrate the effectiveness of the simplified identification method (i.e, stabilized PEM) through a numerical example.
The example shown here is based on a linearized model of the magnetic levitation system (see Fig.~\ref{fig:maglev}) and $H_\infty$ controller given in \cite{Sugie:1993}.  
The system $P$ and controller $K$ are described by 
  \begin{align}
   P(p,\bm{\theta}) &=   
   \frac{\theta_1}{p^3+\theta_2 p^2 +\theta_3 p +\theta_4 }\\
   \bm{\theta} &= [\theta_1, \theta_2, \theta_3, \theta_4 ]^{\top}\\
   &= [-7.148, 13.34, -494.4, -6593]^{\top}\\
   K(p) &=\frac{-1.197 \times 10^8(p+9.294)(p+13.99)(p+20.9)}{(p+399.9)(p+0.1)\left((p+121.5)^2+141.1^2\right)}, \label{eq:maglev_K}
  \end{align}
  where $p$ denotes the differential operator.  Since it is known that the mag. lev. system is order of three without any zeros in continuous-time,
  we try to estimate the four parameters $\theta_1\sim\theta_4$. 
  The closed-loop system for data acquisition is described by
  \begin{align}
    y(t) &= P(p,\bm{\theta}) u(t) + 0.2 \times 10^{-3} \cdot w(t)\\
    u(t) &= K(p) \left(r_y(t)-y(t) \right) + 10 \cdot \xi(t)
  \end{align}
 where the input $u(t)$ of $P(p,\bm{\theta})$ is the voltage applied to the
  electromagnet coil; the output $y(t)$ is the displacement of the steel ball; and $r_y(t)$ is the target displacement and excites the system for identification.
  The measurement noise $w(t)$ and the disturbance $\xi(t)$ are sampled from the standard normal distribution at regular sampling intervals $T_s=10^{-4}\,\mathrm{s}$, and are kept for the interval.

\begin{figure}[tbhp]
    \centering
    \includegraphics[scale=1]{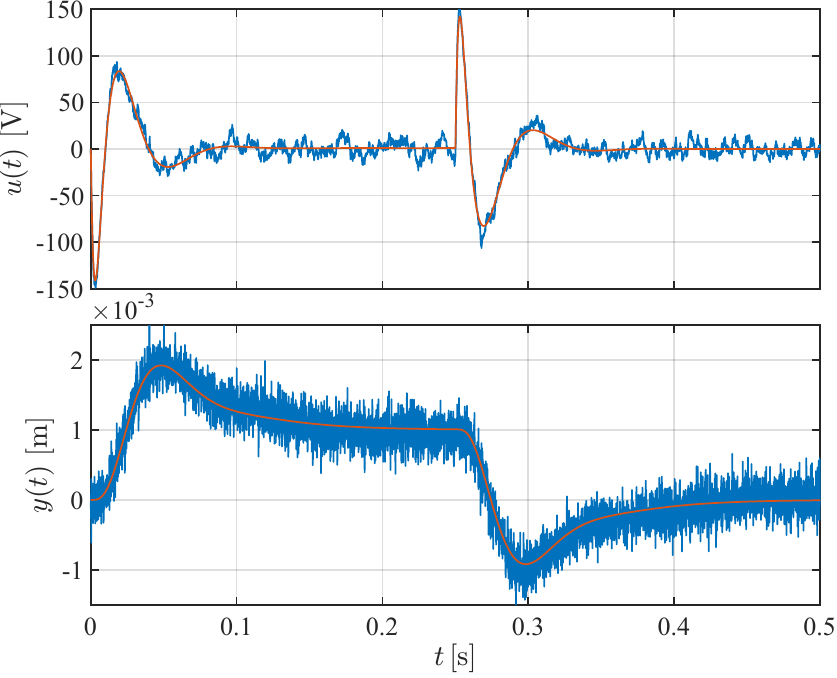}
    \caption{I/O data of magnetic levitation system}
    \label{fig:ex_maglev_io}
  \end{figure}

  An example of I/O data sampled for identification is shown in Fig.~\ref{fig:ex_maglev_io}.
  Here, the target displacement $r_y(t)$ was set to a pulse signal with a width of $0.25\,\mathrm{s}$ and a height of $10^{-3}\,\mathrm{m}$.
  These signals are sampled at intervals $T_s$, and the first-order held signal is used to compute the prediction error $\hat{y}-y$ with $\hat{P} = P(p,\hat{\bm{\theta}})$ and $\hat{K} = K(p)$.
  And, the model $\hat{P}$ is obtained by minimizing the stabilized prediction error (the output of the system in Fig.~\ref{fig:6}) with respect to the parameter estimate $\hat{\bm{\theta}}$.

\begin{figure}[tbhp]
    \centering
    \begin{minipage}{\linewidth}
      \centering
      \includegraphics{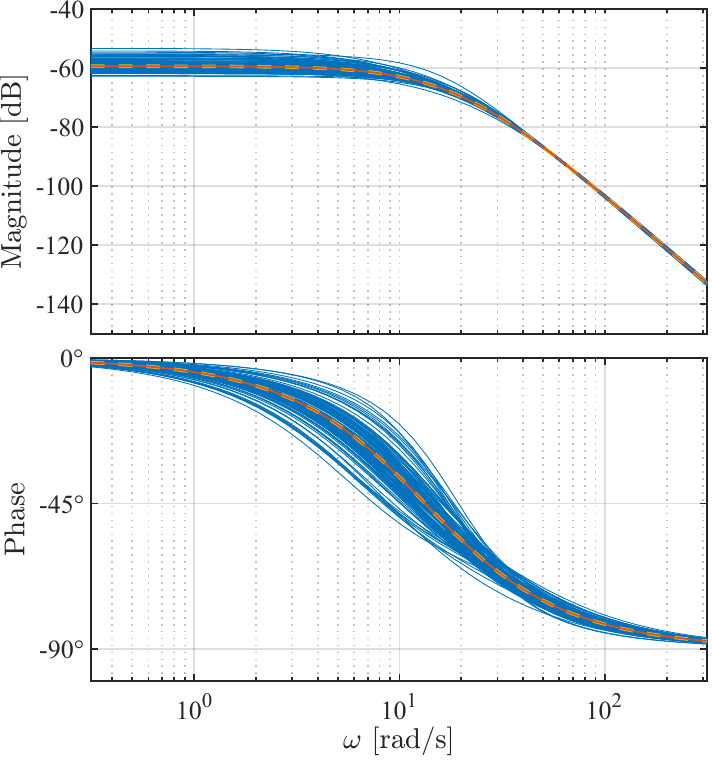}
      \subcaption{With ideal virtual controller ($\hat{K}=K$)}
      \label{fig:ex_maglev_result_soem}
    \end{minipage}
    \begin{minipage}{\linewidth}
      \centering
      \includegraphics{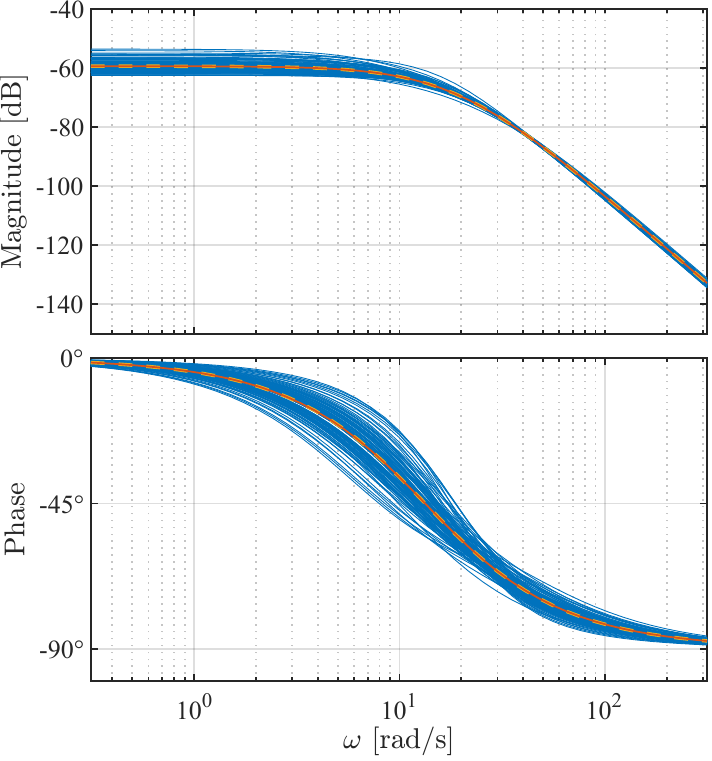}
      \subcaption{With virtual controller different from ideal one ($\hat{K}=K_{\mathrm{PID}} \neq K$)}
      \label{fig:ex_maglev_result_soem_pid}
    \end{minipage}
    \caption{Identification results by the stabilized PEM}
    \label{fig:ex_maglev_result}
  \end{figure}

  Applying the proposed identification method to 100 sets of I/O data,  100 sets of frequency responses of the models are obtained and are shown as blue lines in Fig.~\ref{fig:ex_maglev_result_soem}.
  In the figure, the dashed orange line is the true characteristics of the target system, and the red solid line shows the result obtained from I/O data without disturbance and noise (red lines in Fig.~\ref{fig:ex_maglev_io}).
  To minimize the prediction error about $\hat{\bm{\theta}}$, the Particle Swarm Optimization implementation included in the Global Optimization Toolbox of MATLAB 2020b is used. 

\begin{figure}[tbhp]
    \centering
    \begin{minipage}{\linewidth}
      \centering
      \includegraphics[scale=0.9]{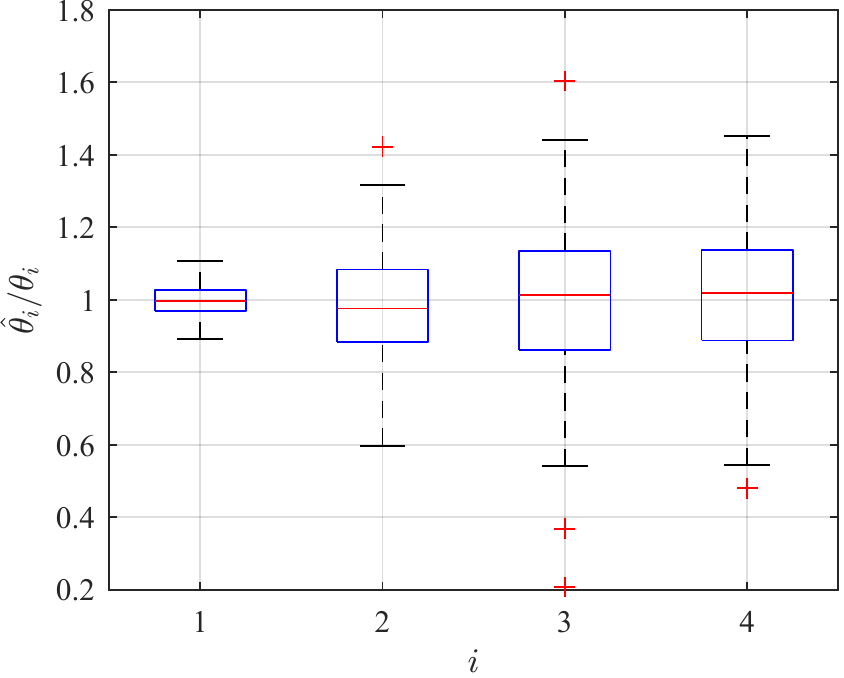}
      \subcaption{With ideal virtual controller ($\hat{K}=K$)}
    \end{minipage}
    \vspace{2mm}
    \begin{minipage}{\linewidth}
      \centering
      \includegraphics[scale=0.9]{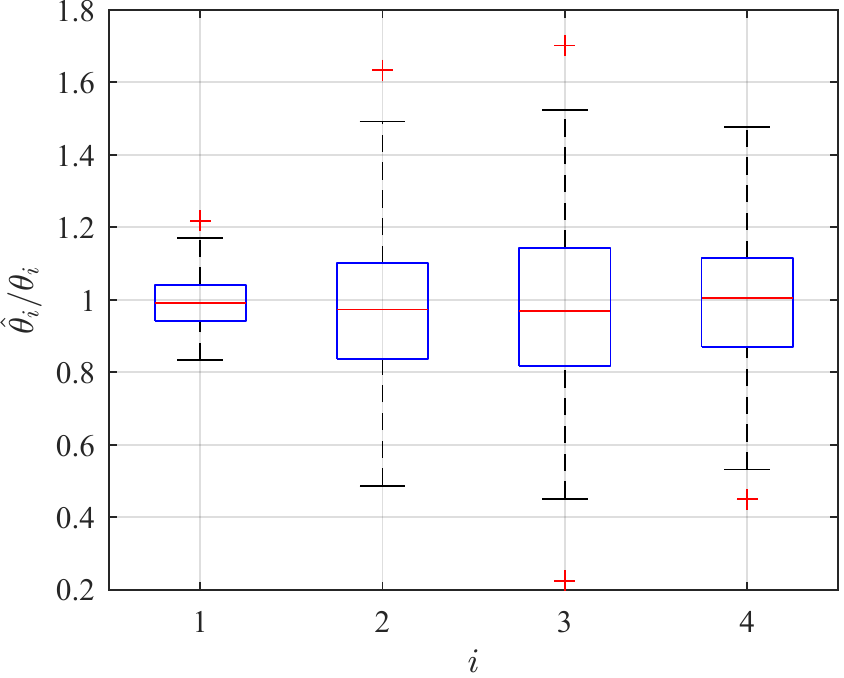}
      \subcaption{With virtual controller different from true one ($\hat{K}=K_{\mathrm{PID}} \neq K$)}
    \end{minipage}
    \caption{Distribution of parameter estimates $\hat{\bm{\theta}}$ by the stabilized PEM}
    \label{fig:ex_maglev_thhat}
  \end{figure}

  To check the sensitivity to the difference between $\hat{K}$ and $K$, the results obtained using the PID controller
  \begin{multline}
    K_{\mathrm{PID}} (p) = \\
    -1798.1\cdot \left( 1 + \frac{1}{0.1438 \cdot p} + \frac{0.1778 \cdot p}{1+8.6336\times10^{-4}p}\right),
  \end{multline}
 which is different from $K$, as virtual controller $\hat{K}$ are shown in Fig.~\ref{fig:ex_maglev_result_soem_pid}.
  Also, the distributions of the obtained estimates are shown in Fig.~\ref{fig:ex_maglev_thhat} as a box plot.
  These results show that the stabilized PEM produces a fairly good model, and there is no significant bias in the parameter estimates, as seen in Fig.~\ref{fig:ex_maglev_thhat}.

  For comparison, we also apply a direct method with well-established (discrete-time) models
  (with sampling period $T_s$), that is,\\
  ARX model:
  \begin{align}
    y(kT_s) &= \frac{B(q)}{A(q)}u(kT_s) + \frac{1}{A(q)} \epsilon(kT_s),
  \end{align}
  ARMAX model:
  \begin{align}
    y(kT_s) &= \frac{B(q)}{A(q)}u(kT_s) + \frac{C(q)}{A(q)} \epsilon(kT_s),
  \end{align}
  Here, 
  \begin{align}
    A(q) &= 1 + a_1 q^{-1} + a_2 q^{-2} + \cdots + a_{n}q^{-n}\\
    B(q) &= b_1 q^{-1} + b_2 q^{-2} + \cdots + b_{n}q^{-n}\\
    C(q) &= 1 + c_1 q^{-1} + c_2 q^{-2} + \cdots + c_{n}q^{-n}   
  \end{align}
  are polynomials of the unit time delay operator $q^{-1}$, and $n$ is the order of the polynomials. 
  The model parameters $a_1$, $\ldots$, $a_{n}$, $b_1$, $\ldots$, $b_{n}$, and $c_1$, $\ldots$, $c_{n}$ 
  are obtained by minimizing the prediction error $\sum_k \epsilon(kTs)^2$.
  The order of each model $n$ is determined based on AIC. 
  For the implementation of the conventional methods, we used the ones provided in System Identification Toolbox of MATLAB 2020b with the default settings.
  
\begin{figure}[thbp]
  \centering
     \begin{minipage}{\linewidth}
        \centering        
        \includegraphics[scale=1]{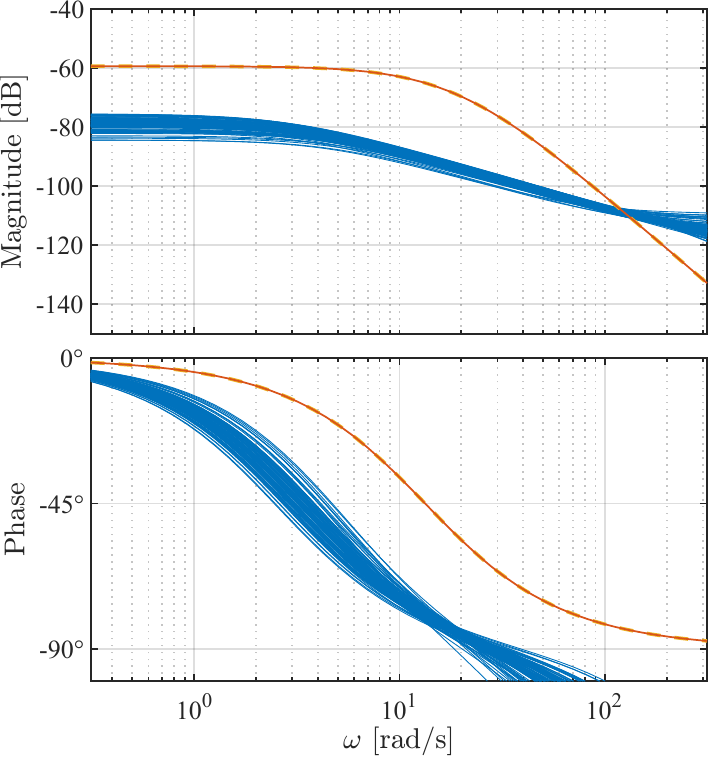} 
      \subcaption{ARX}
    \end{minipage}
\begin{minipage}{\linewidth}     
 \centering
 \includegraphics[scale=1]{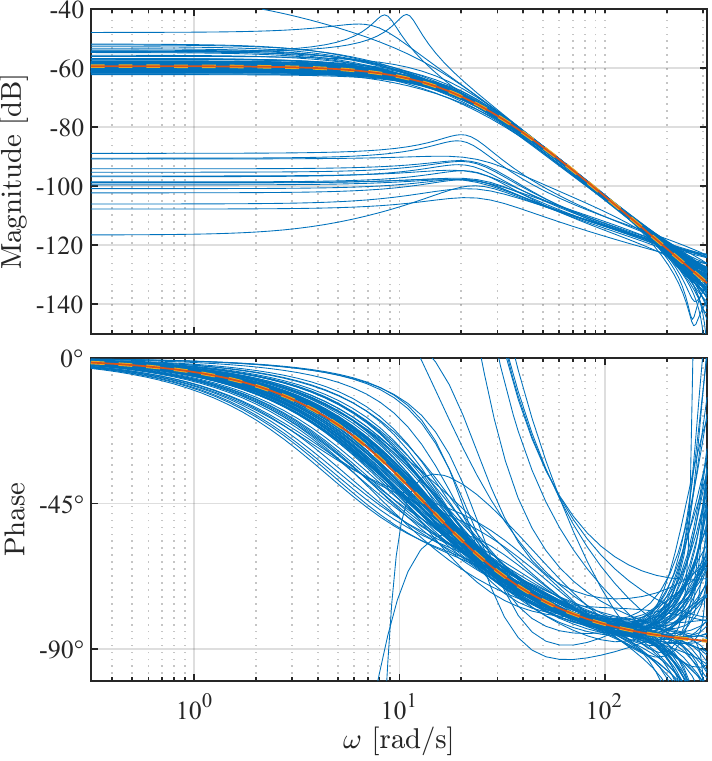}
      \subcaption{ARMAX}
    \end{minipage}%
    \caption{Results of the conventional direct methods}
    \label{fig:ex_maglev_result_conventional}
  \end{figure}

\begin{figure}[bhtp]
    \centering
    \begin{minipage}[b]{\linewidth}
    \centering
    \includegraphics[scale=1]{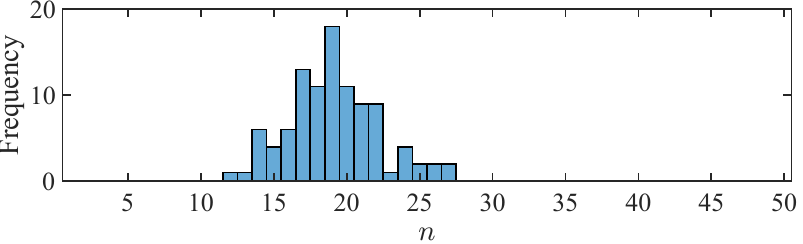}
      \subcaption{ARX}
    \end{minipage}
      \begin{minipage}[b]{\linewidth}
      \centering
      \includegraphics[scale=1]{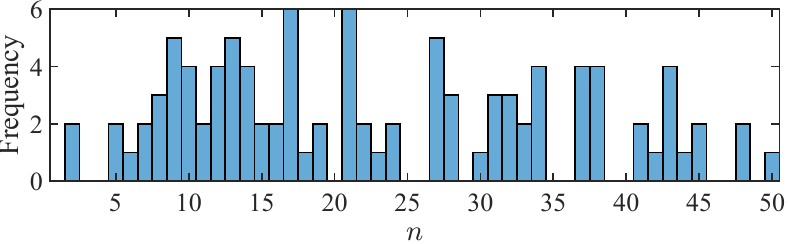}
      \subcaption{ARMAX}
    \end{minipage}
      \caption{Histogram of the selected orders}
    \label{fig:ex_maglev_model_selection}
  \end{figure}

  Here, one hundred identification attempts were made for the I/O data that is identical to that for the stabilized PEM, and the frequency response of the transfer function from $u$ to $y$ of each estimated model is shown as a blue line in Fig.~\ref{fig:ex_maglev_result_conventional}.
  The dashed orange line in the figure is the true characteristics of the target system, and the red solid line shows the result obtained from I/O data without disturbance and noise (red lines in Fig.~\ref{fig:ex_maglev_io}).
  As for the order of the model $n$, the one with the smallest AIC is selected for each data, and the histogram of the selected order is shown in Fig.~\ref{fig:ex_maglev_model_selection}.
  The results show that the stabilized PEM produces better models than the direct methods, 
  which can not use the information on the stabilizing controller.

 In practical applications, one often wants to obtain a good model by performing system identification based on a gray box model that reflects prior information of the target system.
 We next confirm the effectiveness of the stabilized PEM in such case.

 In this example, based on the physical structure of the magnetic levitation system, we can also consider a gray box model of the following structure:
  \begin{align}
    P(p,\bm{\theta}) &= \frac{-K_i}{\left(Mp^2-K_x\right)\left(Lp+R\right)}\\
    \bm{\theta} &= \left[K_i, K_x\right]^\top
  \end{align}
  Here, $R=27.03\,\mathrm{\Omega}$ and $L=2.027\,\mathrm{H}$ are the resistance and inductance of the electromagnet, respectively; $M=0.358\,\mathrm{kg}$ is mass of the ball; $K_i = 5.187\,\mathrm{N/A}$ and $K_x=177.0\,\mathrm{N/m}$ are constants related to the electromagnetic force at the equilibrium point.
  Since direct measurements are possible for $R$, $L$ and $M$, our problem here is to estimate $K_i$ and $K_x$ while assuming $R$, $L$ and $M$ are given.

\begin{figure}[hbtp]
    \centering
    \begin{minipage}{\linewidth}
      \centering
      \includegraphics{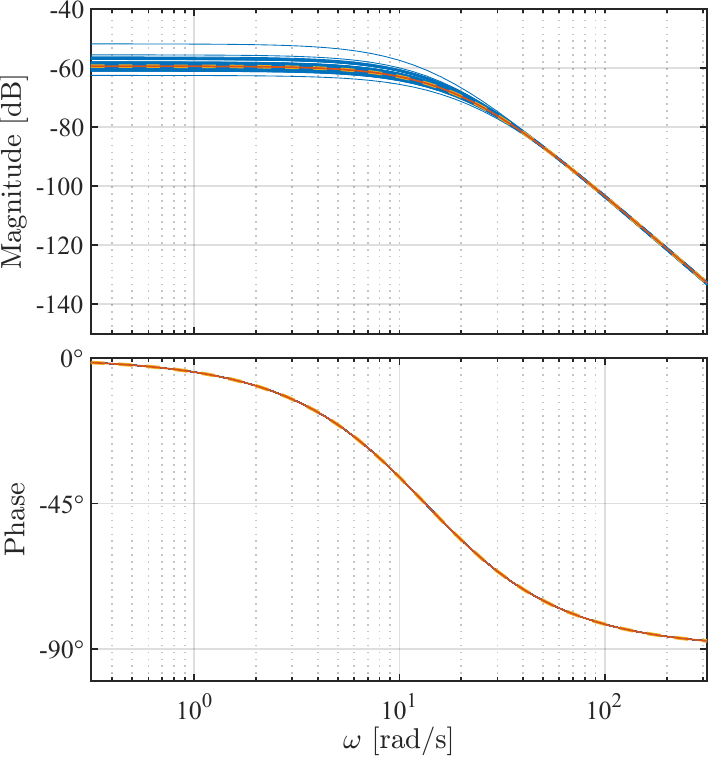}
      \subcaption{With ideal virtual controller ($\hat{K}=K$)}
      \label{fig:ex_maglev_gray_result_soem}
    \end{minipage}
    \begin{minipage}{\linewidth}
      \centering
      \includegraphics{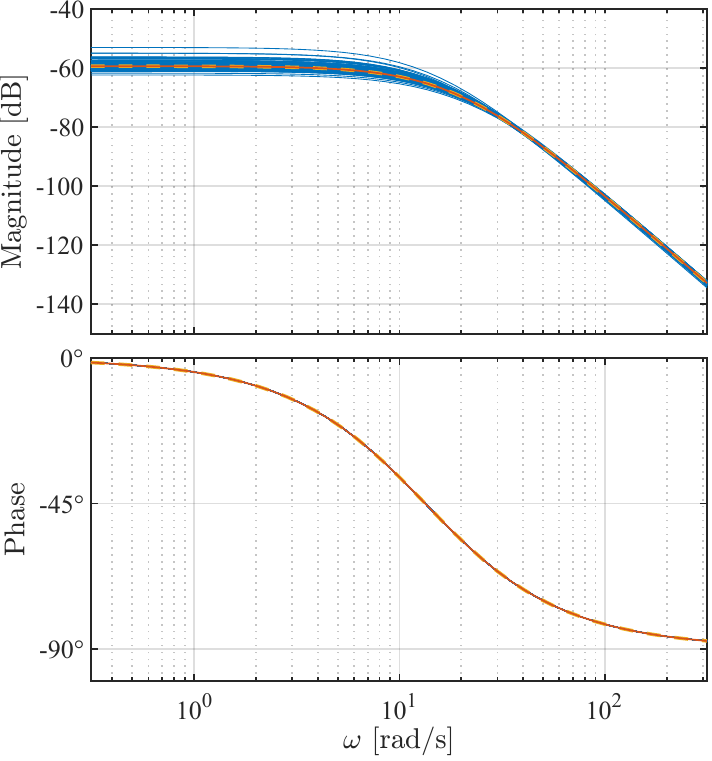}
      \subcaption{With virtual controller different from ideal one ($\hat{K}=K_{\mathrm{PID}} \neq K$)}
      \label{fig:ex_maglev_gray_result_soem_pid}
    \end{minipage}
    \caption{Identification results by the stabilized PEM with gray box model}
    \label{fig:ex_maglev_gray_result}
  \end{figure}

\begin{figure}[hbtp]
    \centering
    \begin{minipage}[t]{0.49\linewidth}
      \centering
      \includegraphics[scale=0.9]{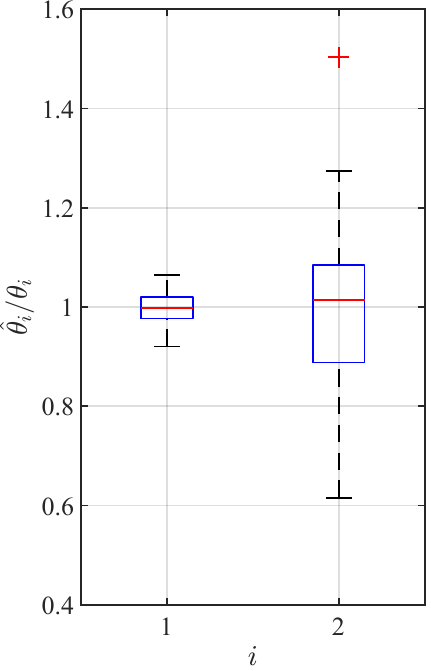}
      \subcaption{With ideal virtual controller ($\hat{K}=K$)}
    \end{minipage}
    \begin{minipage}[t]{0.49\linewidth}
      \centering
      \includegraphics[scale=0.9]{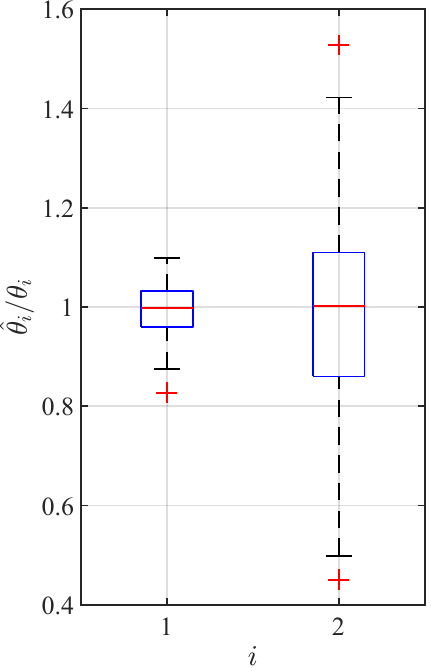}
      \subcaption{With virtual controller different from true one ($\hat{K} =K_{\mathrm{PID}}\neq K$)}
    \end{minipage}
    \caption{Distribution of parameter estimates $\hat{\bm{\theta}}$ by the stabilized PEM}
    \label{fig:ex_maglev_gray_thhat}
  \end{figure}

  The results obtained based on this gray box model from the aforementioned data are shown in Figs.~\ref{fig:ex_maglev_gray_result} and \ref{fig:ex_maglev_gray_thhat}.
  As can be seen by comparing Figs.~\ref{fig:ex_maglev_result} and \ref{fig:ex_maglev_gray_result}, the variance of the estimates is effectively reduced by using the known information about the structure.
  Also, the results are not much affected by the choice of the virtual controller $\hat{K}$.


 \section{Conclusion}
This paper has shown how to derive a simplified version of the dual Youla method which identifies the plant $P$ itself (instead of Youla parameter $Q$) without coprime factorization, while enjoying all the merits (such as robustness against both controller parameter uncertainty and noise model structure, and easiness to handle unstable plants) of the dual Youla method. This simplified version turned out to be identical to the stabilized PEM which was proposed by the authors recently. Since coprime factorization nor any special identification technique is not necessary at all, it is easy for most engineers (i.e., non-experts on identification) to use it with numerical optimization. Detailed simulation results have been given to demonstrate the above merits.



                                                   







\end{document}